\documentclass[
superscriptaddress,
twocolumn,
showpacs,preprintnumbers,
amsmath,amssymb,
aps,
prb,
]{revtex4-2}
\usepackage[colorlinks,
linkcolor=blue, urlcolor=blue, anchorcolor=blue, citecolor=blue]{hyperref}
\usepackage{graphicx}
\usepackage{dcolumn}
\usepackage{bm}
\usepackage{float}
\usepackage{multirow}
\usepackage{titlesec}
\usepackage{xcolor}
\usepackage{lipsum}  
\usepackage[normalem]{ulem}
\usepackage{balance} 

\begin{document}


\title{Multiferroic Metallic Monolayer Cu(CrSe$_2$)$_2$} 


\author{Ke Yang}
\thanks{K. Y. and Y. Z. contributed equally to this work.}
\affiliation{College of Science, University of Shanghai for Science and Technology, Shanghai 200093, China}
 \affiliation{Laboratory for Computational Physical Sciences (MOE),
 State Key Laboratory of Surface Physics, and Department of Physics,
 Fudan University, Shanghai 200433, China}

 \author{Yuxuan Zhou}
 \thanks{K. Y. and Y. Z. contributed equally to this work.}
 \affiliation{Laboratory for Computational Physical Sciences (MOE),
 State Key Laboratory of Surface Physics, and Department of Physics,
 Fudan University, Shanghai 200433, China}

 \author{Yaozhenghang Ma}
 \affiliation{Laboratory for Computational Physical Sciences (MOE),
 State Key Laboratory of Surface Physics, and Department of Physics,
 Fudan University, Shanghai 200433, China}

 \author{Hua Wu}
 \email{Corresponding author. wuh@fudan.edu.cn}
 \affiliation{Laboratory for Computational Physical Sciences (MOE),
  State Key Laboratory of Surface Physics, and Department of Physics,
  Fudan University, Shanghai 200433, China}
 \affiliation{Shanghai Qi Zhi Institute, Shanghai 200232, China}
 \affiliation{Hefei National Laboratory, Hefei 230088, China}

\date{\today}

\begin{abstract}
  The two-dimensional (2D) Cu(CrSe$_2$)$_2$ monolayer stands out for its combined ferromagnetic (FM), ferroelectric (FE), and metallic properties, marking itself as a prominent 2D multiferroic metal.
  This work studies those properties and the relevant physics, using density functional calculations, Monte Carlo simulations, and $ab$ $initio$ molecular dynamics.
  Our results show that Cu(CrSe$_2$)$_2$ monolayer is in the Cr$^{3+}$ $t_{2g}^3$ state with $S$ = 3/2 and Cu$^{1+}$ $3d^{10}$ with $S$ = 0. 
  A ligand hole in the Se 4$p$ orbitals gives rise to metallic behavior and enhances the FM coupling between the local Cr$^{3+}$ $S$ = 3/2 spins.
  The observed in-plane magnetic anisotropy primarily arises from exchange anisotropy, which is associated with the Cr-Se-Cr itinerant ferromagnetism. In contrast, both single-ion anisotropy and shape magnetic anisotropy contribute negligibly. The Dzyaloshinskii-Moriya interaction is also quite weak, only about 3\% of the intralayer exchange parameters. Our Monte Carlo simulations show a FM Curie temperature ($T_{\rm C}$) of 190 K. Moreover, the monolayer exhibits a vertical FE polarization of 1.79 pC/m and a FE polarization switching barrier of 182 meV/f.u., and the FE state remains stable above room temperature as shown by $ab$ $initio$ molecular dynamics simulations.
  Furthermore, a magnetoelectric coupling is partially manifested by a magnetization rotation from in-plane to out-of-plane associated with a FE-to-paraelectric transition. The magnetization rotation can also be induced by either hole or electron doping, and the hole doping increases the $T_{\rm C}$ up to 238 K.
  In addition, tensile strain reduces the FE polarization but enhances $T_{\rm C}$ to 290 K, while a compressive strain gives an opposite effect. Therefore, the multiferroic metallic Cu(CrSe$_2$)$_2$ monolayer may be explored for advanced multifunctional electronic devices.
\end{abstract}
\maketitle


Research on two-dimensional (2D) materials has substantially expanded due to the discovery of 2D ferromagnetic (FM) materials, including CrI$_3$~\cite{Huang2017} and Cr$_2$Ge$_2$Te$_6$~\cite{Gong2017}, and 2D ferroelectric (FE) materials, such as CuInP$_2$S$_6$~\cite{liu2016} and SnTe~\cite{chang2016}.
The Mermin-Wagner theorem posits that long-range magnetic order cannot be sustained at any finite temperature in 2D isotropic Heisenberg spin systems~\cite{MW}. 
This limitation of 2D systems challenges the conventional understanding of magnetic behavior and underscores the importance of magnetic anisotropy (MA) in stabilizing their magnetic states~\cite{Burch_2018,gibertini2019}.
Ferroelectricity, a characteristic of certain insulators, displays spontaneous electric polarization reversible by an external electric field~\cite{dawber2005,scott2007}.
Anderson and Blount introduced the concept of ``FE metal"~\cite{Anderson1965}. During a symmetry-breaking structural transition, it is hypothesized that a metal could generate FE. This hypothesis has been recently confirmed through 3D LiOsO$_3$ and NdNiO$_3$~\cite{shi2013,kim2016}.
In the realm of 2D structures, FE metals are exceedingly rare~\cite{wang2023}. To date, the only evidence is the demonstrated ability to reverse polarization in a 2D WTe$_2$ FE metal by applying an electric field~\cite{Fei2018}.
When magnetization and FE metallicity are combined in the 2D limit, the result is a ``2D multiferroic metal". This material not only exhibits characteristics typical of bulk multiferroics, such as phase transitions~\cite{li2021}, magneto-electric coupling~\cite{song2022}, and unconventional optical responses~\cite{kurebayashi2022}, but also offers unique advantages due to its monolayer thickness. These properties can be effectively manipulated using magnetic and electric fields, doping, Moir$\acute{\rm {e}}$ engineering, or the creation of heterostructures~\cite{gibertini2019,xu2022,mak2019,song_2018,Song2019,Li2019,Saez2023,castro2024}.

\begin{figure}[t]
  \centering
 \includegraphics[width=8.5cm]{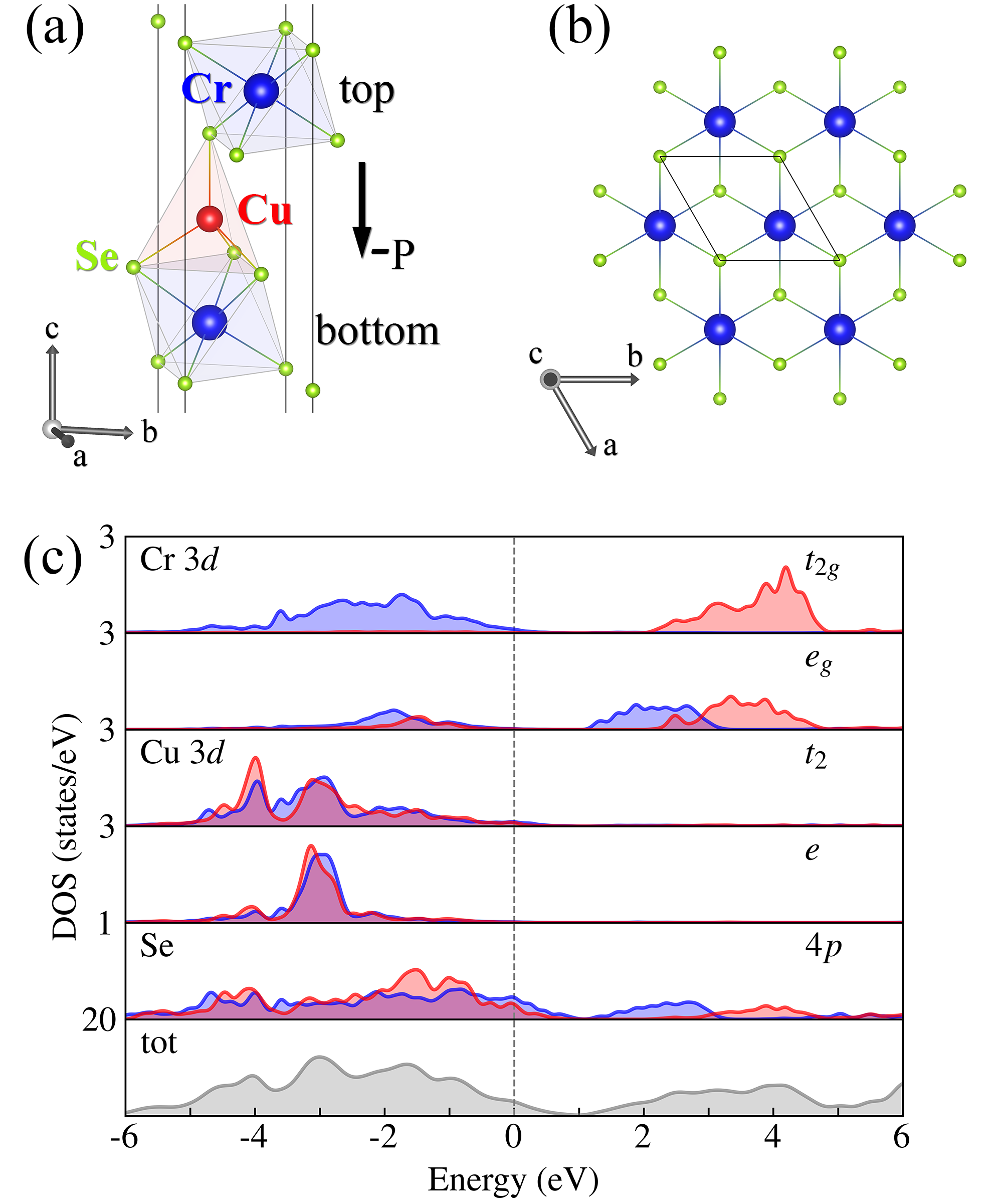}
 \centering
 \caption{(a) Side view and (b) top view of the Cu(CrSe$_{2}$)$_{2}$ monolayer. In (b), only the top CrSe$_{2}$ layer is depicted. (c) The Cr 3$d$, Cu 3$d$, Se 4$p$ and total DOS by GGA+SOC+$U$. The blue (red) curves stand for the up (down) spin channel. The Fermi level is set at zero energy.
}
 \label{struct}
 \end{figure}

Very recently, the 2D Cu(CrSe$_2$)$_2$ monolayer was derived from CuCrSe$_2$ bulk~\cite{Cirish2014,niedziela2019} using a redox-controlled electrochemical exfoliation method~\cite{Peng2023,sun2024}. As depicted in \textbf{Figure~\ref{struct}}, this monolayer features a Cu atom layer sandwiched between two CrSe$_2$ layers, where Cr-cations form triangular nets through edge-shared CrSe$_6$ octahedra. Each Cu atom forms a tetrahedral bond, connecting to one Se atom from the top CrSe$_2$ layer and three Se atoms from the bottom layer. Experimental investigations have confirmed that the Cu(CrSe$_2$)$_2$ monolayer exhibits FM behavior with a Curie temperature (\(T_{\mathrm{C}}\)) of 125 K and notable in-plane magnetization~\cite{Peng2023}. Meanwhile, the FE phase transition temperature for ultrathin CuCrSe$_2$ has been measured at approximately 800 K~\cite{Wang_2024}. Previously, the multiferroic behavior was also theoretically predicted to be above 300 K~\cite{Zhong2019}, which is in contrast with the experimental $T_{\rm C}$ of 125 K. More recently, a study has reported that single-layer Cu(CrSe$_2$)$_2$ not only maintains ferroelectricity above room temperature but also displays ferromagnetism at 120 K~\cite{sun2024}, further underscoring its intriguing multiferroic nature.

Here, we are motivated to study those multiferroic properties of Cu(CrSe$_2$)$_2$ monolayer and particularly the relevant physics, using density functional calculations, Monte Carlo (MC) simulations, and $ab$ $initio$ molecular dynamics. Our results show that the itinerant Se $4p$ holes enhance the Cr-Cr FM coupling and the exchange anisotropy determines the observed in-plane MA. Our MC simulations give the FM $T_{\rm C}$ of 190 K, which is moderately higher than the experimental 120-125 K~\cite{Peng2023,sun2024}.
Moreover, the monolayer exhibits a vertical FE polarization of 1.79 pC/m and a FE switching barrier of 182 meV/f.u., and the FE state remains stable above room temperature as proven by our $ab$ $initio$ molecular dynamics simulations. Furthermore, we predict that the magnetization would rotate from in-plane to out-of-plane upon a FE-to-paraelectric transition, showing a magnetoelectric coupling. The magnetization rotation can also be induced by either hole or electron doping, and a hole doping would increase $T_{\rm C}$ up to 238 K. A tensile strain would reduce the FE polarization but elevate $T_{\rm C}$ up to 290 K. 
Thus, our work provides a systematic and comprehensive investigation of the multiferroic properties of Cu(CrSe$_2$)$_2$ monolayer, potentially aiding in the exploration of its multifunctionality in spintronics.

\section*{Computational Details}

We perform density functional theory (DFT) calculations using the Vienna \textit{ab} $initio$ simulation package (VASP)~\cite{VASP}.
The kinetic energy cutoff for the plane wave expansion is set to 500 eV.
The $\varGamma$-centered mesh of 15$\times$15 is used for the 1$\times$1 unit cell and (15$\times$8) k-mesh for the 1$\times$2 supercell. 
The setup of kinetic energy and k-point sampling are carefully tested to ensure the convergence of the MA results within a few percent, as shown in Table S1 in supporting information (SI).
Atomic positions are fully relaxed until the force on each atom is less than 0.01 eV/\AA~, and the total energy minimization is performed with a tolerance of 10$^{-7}$ eV.
The DFT-optimized lattice constants $a$ = $b$ = 3.63 $\AA$ with space group $P3m1$ are used for the Cu(CrSe$_2$)$_2$ monolayer, which are almost the same as the experimental bulk values of $a$ = $b$ = 3.69 $\AA$~\cite{Cirish2014}.
A vacuum space larger than 15 \AA~ is used.

To account for the electron correlation of the Cu (Cr) $3d$ states, the generalized gradient approximation plus Hubbard 
$U$ (GGA + $U$)~\cite{Vladimir_1997} calculations are performed with Hubbard $U$ = 6.0 eV (4.0 eV) and Hund exchange $J_{\rm H}$ = 0.9 eV (0.9 eV). 
The Hubbard $U$ was also used in the previous calculations of the FM metal CrO$_2~$\cite{CrO2PRL}. The SOC is included for Cu, Cr, and Se atoms by the second-variation method with scalar relativistic wave functions.
We also tested \(U_{\rm {Cu}} = 7\) eV and \(U_{\rm {Cr}} = 5\) eV (see Table S2 in the SI). Although increasing \(U\) results in some quantitative changes in the exchange parameters and the magnetic anisotropy energy (MAE), our conclusions remain unaffected, $e.g.$, the estimated \(T_{\mathrm{C}}\) of the Cu(CrSe$_2$)$_2$ monolayer undergoes only a small change from 190 K to 178 K. 
Note that if the Hubbard $U$ is not used, the plain GGA calculations (see Fig. S1) overestimate the itineracy of the Cr $3d$ electrons and give the enhanced FM exchange and exchange anisotropy (see Table S2), and then the MC simulation gives $T_{\rm C}$ of 360 K, being much higher than the experimental 120-125 K~\cite{Peng2023,sun2024}.
Dipole moment correction is applied to evaluate the vertical polarization and the FE switching path is calculated using the climbing image nudged elastic band (NEB) method~\cite{CINEB}.
The phonon spectrum is calculated using Phonopy software interfaced with VASP~\cite{phonopy}.
In addition, $ab$ $initio$ molecular dynamics (AIMD) simulations are performed to study the stability of the FE polarization in a 4$\times$4 supercell over 30 ps at different temperatures. The magnetic phase transition of the Cu(CrSe$_2$)$_2$ monolayer is studied using MC simulations.
In our MC simulations, a $15\times15\times1$ lattice is chosen. During the simulation step, each spin is randomly rotated in three-dimensional space. The spin dynamical process is studied by the classical Metropolis methods~\cite{Metropolis}.

\section*{Results and Discussion}

\subsection*{Electronic states and magnetic properties}
First, we investigate the charge state and the metallic nature of the Cu(CrSe$_2$)$_2$ monolayer using the GGA+SOC+$U$ calculations. 
We plot in Fig.~\ref{struct}(c) the orbitally resolved density of states (DOS) for the FM state. 
Within each CrSe$_2$ layer, six Se ions octahedrally coordinate the Cr ion, which causes the $d$ orbitals to split into low $t_{2g}$ and high $e_g$ levels. The up-spin Cr $t_{2g}$ states are fully occupied, while the up-spin $e_g$ states are primarily unoccupied, leading to a formal Cr$^{3+}$ $t_{2g}^3$ $S$ = 3/2 configuration. The marginally occupied $e_g$ states below the Fermi level arise from strong Cr $e_g$-Se 4$p$ $pd\sigma$ hybridization.
Moreover, the Cr$^{3+}$ $S$ = 3/2 charge-spin state is confirmed by localized spin moments of 3.04 and 3.12 $\mu_{\rm B}$ in the top and bottom CrSe$_2$ layers, respectively.
In the local tetrahedral crystal field, the splitting of the $e$ and $t_2$ states for Cu is small. Our DOS results confirm this minimal crystal field splitting, as seen in Fig.~\ref{struct}(c). Both the $e$ and $t_2$ states are fully occupied, indicating a formal Cu$^{1+}$ 3$d^{10}$ $S = 0$ configuration.
The tiny spin moment of Cu ions, --0.05 $\mu_{\rm B}$, corresponds to the nominal state of Cu$^{1+}$ with $S$ = 0 again.
As a result, the Cu(CrSe$_2$)$_2$ monolayer is characterized by a charge configuration of Cr$^{3+}$ and Cu$^{1+}$. 
In line with this configuration, the Se ions are anticipated to have a hole per formula unit (f.u.) to maintain electrical neutrality.
This is evidenced by the Fermi level traversing Se's 4$p$ bands and leaving an unoccupied hole, as illustrated in Fig.~\ref{struct}(c).
The negative spin moment of $-$0.16 $\mu_{\rm B}$ on each Se ligand is induced by its adjacent Cr$^{3+}$ ions with $S$ = 3/2. 
Consequently, the ligand holes in the Se 4$p$ bands determine the observed metallic behavior of Cu(CrSe$_2$)$_2$, and the itinerant Se 4$p$ holes would enhance, via the Cr $3d$- Se $4p$ hybridizations, the FM coupling between the localized Cr $S$ = 3/2 spins.

\begin{figure}[t]
  \centering
 \includegraphics[width=8cm]{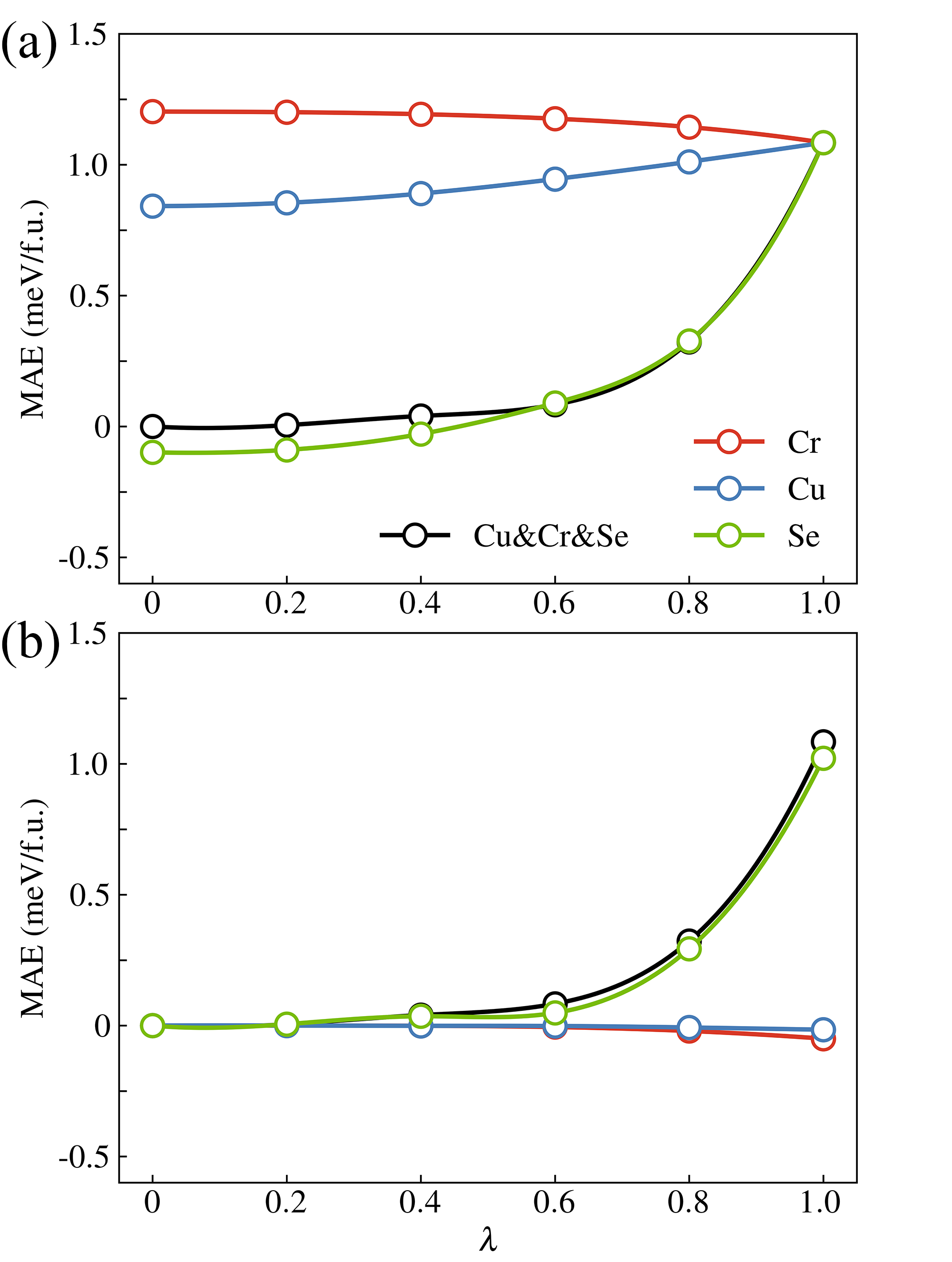}
 \centering
 \caption{Black line: evolution of the MAE as a function of $\lambda_{\rm {Cu, Cr, Se}}$. (a) Red, blue, and green lines: evolution of the MAE as a function of $\lambda_{\rm{Cr}}$, $\lambda_{\rm {Cu}}$, and $\lambda_{\rm {Se}}$, while keeping two others at 1. (b) Red, blue, and green lines: evolution of the MAE as a function of $\lambda_{\rm{Cr}}$, $\lambda_{\rm {Cu}}$, and $\lambda_{\rm {Se}}$, while keeping two others at 0.
 }
 \label{MAE}
 \end{figure}

MA is extremely important for the stability of long-range magnetic order in 2D magnetic materials. Next, we will investigate the easy magnetization direction of the Cu(CrSe$_2$)$_2$ monolayer.
The inherent $C_3$ symmetry of the material along the $c$-axis suggests the in-plane isotropy, and thus one just needs to compare the in-plane and out-of-plane magnetizations to determine the easy magnetization axis.
Our results indicate that the Cu(CrSe$_2$)$_2$ monolayer favors in-plane magnetization, in line with experimental observations~\cite{Peng2023}, while the out-of-plane magnetization is higher in energy by 1.08 meV/f.u..

MA has several significant sources in 2D magnets, such as single-ion anisotropy (SIA), magnetic dipole-dipole interaction (shape-MA), and exchange anisotropy (EA)~\cite{Lado2017,Kim_2019,Yang2020,liu2020,Yang2021}. Here, we identify the primary source of in-plane MA in the Cu(CrSe$_2$)$_2$ monolayer.
The SIA results from the intrinsic spin-orbit coupling effect $\lambda\overrightarrow{L}\cdot\overrightarrow{S}$, which requires the presence of the orbital moment. In Cu(CrSe$_2$)$_2$, the magnetic Cr$^{3+}$ ions have a closed $t_{2g}^3$ shell, seemingly precluding any orbital degrees of freedom.
Consequently, these Cr$^{3+}$ ions display a nearly zero orbital moment of --0.001 $\mu_{\rm B}$, leading to a negligible SIA.
To further confirm this, we adjust the strength of $\lambda_{\rm Cr}$ to values different from the natural spin-orbit coupling strength, while maintaining the inherent spin-orbit coupling strengths of Cu and Se at their natural level of 1 (see Fig.~\ref{MAE}(a)). Despite these adjustments, the MAE of the material remains unchanged.
Conversely, when the strength of $\lambda_{\rm Cr}$ is adjusted with Cu and Se at 0 (see Fig. ~\ref{MAE}(b)), the material's MAE becomes zero. This suggests that in Cu(CrSe$_2$)$_2$, the contribution of SIA from magnetic ion Cr$^{3+}$ is essentially negligible.
For Cr$^{3+}$ with FM coupling, the shape-MA favors in-plane orientation and manifests at 165 $\mu$eV/f.u., suggesting that it is not the dominant factor in this material~\cite{Yang2021}. Additionally, we consider the role of EA, which arises from the Se $4p$-Cr $3d$ exchange mediating the FM Cr-Cr coupling via the itinerant holes in the heavy Se 4$p$ orbital.
As seen in Fig. ~\ref{MAE}, the strength of $\lambda_{\rm Se}$ determines the overall trend of the material's MAE. If $\lambda_{\rm Se}$ is zero, the material's MAE is almost zero. However, when $\lambda_{\rm Se}$ is 1 (its natural SOC strength), the material exhibits easy in-plane magnetization, being almost independent of $\lambda_{\rm Cu}$ and $\lambda_{\rm Cr}$.
Thus, the easy in-plane MA of Cu(CrSe$_2$)$_2$ predominantly arises from EA, while SIA and shape-MA are negligible. This is similar to the previous proposal that EA is the main source of MA in CrI$_3$~\cite{Lado2017,Kim_2019}.

With the confirmed easy in-plane magnetization, we now turn to the estimation of the magnetic interactions. The Cu(CrSe$_2$)$_2$ monolayer is characterized by Cr$^{3+}$ $S$ = 3/2 and Cu$^{1+}$ $S = 0$ charge states, with a metallic ligand hole predominantly on the Se 4$p$ band. One thus may expect an itinerant FM ground state.
To substantiate this, we employed GGA+SOC+$U$ calculations on a 1$\times$2 supercell, examining three additional antiferromagnetic (AF) states with in-plane magnetization, as depicted in Fig. S2 in SI.
Indeed, the FM metallic state is more stable than the three AF states, as seen in Table S3 in SI. Our results indicate that both CrSe$_2$ layers and the Cu-connected interlayer exhibit FM coupling.
In order to obtain the exchange parameters, we introduced three specific ones: $J_{\mathrm{top}}$ for the top layer of CrSe$_2$, $J_{\mathrm{bottom}}$ for the bottom layer, and $J_{\mathrm{inter}}$ for the interactions between the CrSe$_2$ layers, as illustrated in Fig. S2 in SI. Considering that the nearest-neighbor Cr ions in the CrSe$_2$ layers are six-coordinated and those in the interlayer are three-coordinated, we express as follows the magnetic exchange energies for the FM ground state and three AF states per formula unit, with an assumption of $JS^2$ for a FM pair of Cr$^{3+}$ $S$ = 3/2 ions:

\begin{equation}\begin{split}
E_{\rm {FM}}&=(\ 3J_{\rm {top}}+3J_{\rm {bottom}}+3J_{\rm {inter}})S^2+E_0\\
E_{\rm {AF1}}&=(-J_{\rm {top}}+3J_{\rm {bottom}})S^2+E_0\\
E_{\rm {AF2}}&=(\ 3J_{\rm {top}}-\ J_{\rm {bottom}})S^2+E_0\\
E_{\rm {inter-AF}}&=(\ 3J_{\rm {top}}+3J_{\rm {bottom}}-3J_{\rm {inter}})S^2+E_0\\
\end{split}\end{equation}
Using the calculated relative total energies (see Table S3 in SI) and the above equations, we derive the intralayer $J_{\rm {top}}$
= $-$5.70 meV and $J_{\rm {bottom}}$ = $-$6.55 meV, and the interlayer $J_{\rm {inter}}$ =$-$0.18 meV being one order of magnitude weaker.
These negative values indicate that both intralayer and interlayer couplings are FM. Note that for the small-gap semiconducting bulk of CuCrSe$_2$ (without the formal Se $4p$ holes), we also calculate its intralayer Cr$^{3+}$-Cr$^{3+}$ FM parameter to be $-$4.33 meV. Therefore, in the Cu(CrSe$_2$)$_2$ monolayer, the itinerant ligand Se $4p$ holes enhance the FM interactions.

The Cu(CrSe$_2$)$_2$ monolayer, with the space group $P$3$m$1 and lacking spatial inversion symmetry, exhibits also Dzyaloshinskii-Moriya (DM) interactions~\cite{Dzyaloshinsky1958,Moriya1960}. We now estimate the DM interactions using the four-state energy mapping approach~\cite{xiang2013}.
As an example, we calculate the interactions between a pair of nearest Cr ions 
along the $b$-axis. The results for the DM interaction are [0.09, --0.08, 0.09] meV, with $\left\lvert D\right\rvert$  = 0.15 meV for the top layer and [0.17, --0.11, 0.00] meV, with $\left\lvert D\right\rvert$  = 0.20 meV for the bottom layer, see Tables S4 and S5 in SI. 
The $\left\lvert D\right\rvert$/$\left\lvert J\right\rvert$ ratios on the top and bottom Cr layers are 2.6\% and 3.0\%, respectively.
The DM interactions are much weaker than the intralayer exchange parameters in this material and can therefore be neglected in this work.

Using the calculated FM parameters and in-plane MA, we propose the following spin Hamiltonian to estimate the $T_{\rm C}$ of the Cu(CrSe$_2$)$_2$ monolayer:

\begin{align}
  \begin{split}
  H = &J_{\rm {top, bottom, inter}}\sum_{\langle{i,j}\rangle}{\overrightarrow{S_{i}} \cdot \overrightarrow{S_{j}}} + J'\sum_{\langle{i,j}\rangle}{S^{z}_{i} \cdot S^{z}_{j}} 
  \label{eq6}
  \end{split}
\end{align}
The sum over $i$ runs over the entire lattice of Cr atoms, and $j$ runs over the six nearest Cr neighbors of each Cr atom. The first term describes the intralayer and interlayer isotropic Heisenberg exchange interaction. The second term represents the exchange anisotropy, which is caused by the SOC effect of the ligand Se $4p$ holes mediating the FM Cr-Cr coupling, giving the energy difference between in-plane and out-of-plane magnetizations. In our calculations, the exchange anisotropy constant $J'$ is determined to be 0.04 meV, and it corresponds to the easy in-plane magnetization.
Our MC simulations show that the Cu(CrSe$_2$)$_2$ monolayer has a $T_{\mathrm{C}}$ of 190 K, which is much lower than the previous theoretical prediction of $T_{\mathrm{C}}$ over 300 K~\cite{Zhong2019} but is a moderate overestimation,   
compared with the experimental values of 120-125 K~\cite{Peng2023,sun2024}.

Considering the FM metallic nature of the Cu(CrSe$_2$)$_2$ monolayer and possible long-range magnetic couplings, we also calculate the exchange parameters up to the third-nearest neighbors ($J_3$), as detailed in Fig. S3, Equation S1, and Table S6 in SI. These results show that although the specific exchange parameters vary upon the inclusion of the different magnetic couplings in the spin Hamiltonian, here the $J_2$ and $J_3$ for both the top and bottom CrSe$_2$ layers are nearly one order of magnitude weaker than the above $J_{\mathrm{top}}$ and $J_{\mathrm{bottom}}$, and a core result--the estimated $T_{\mathrm{C}}$ remains almost unchanged, being 185 K given by our MC simulations.

\begin{figure}[t]
  \centering
 \includegraphics[width=8.5cm]{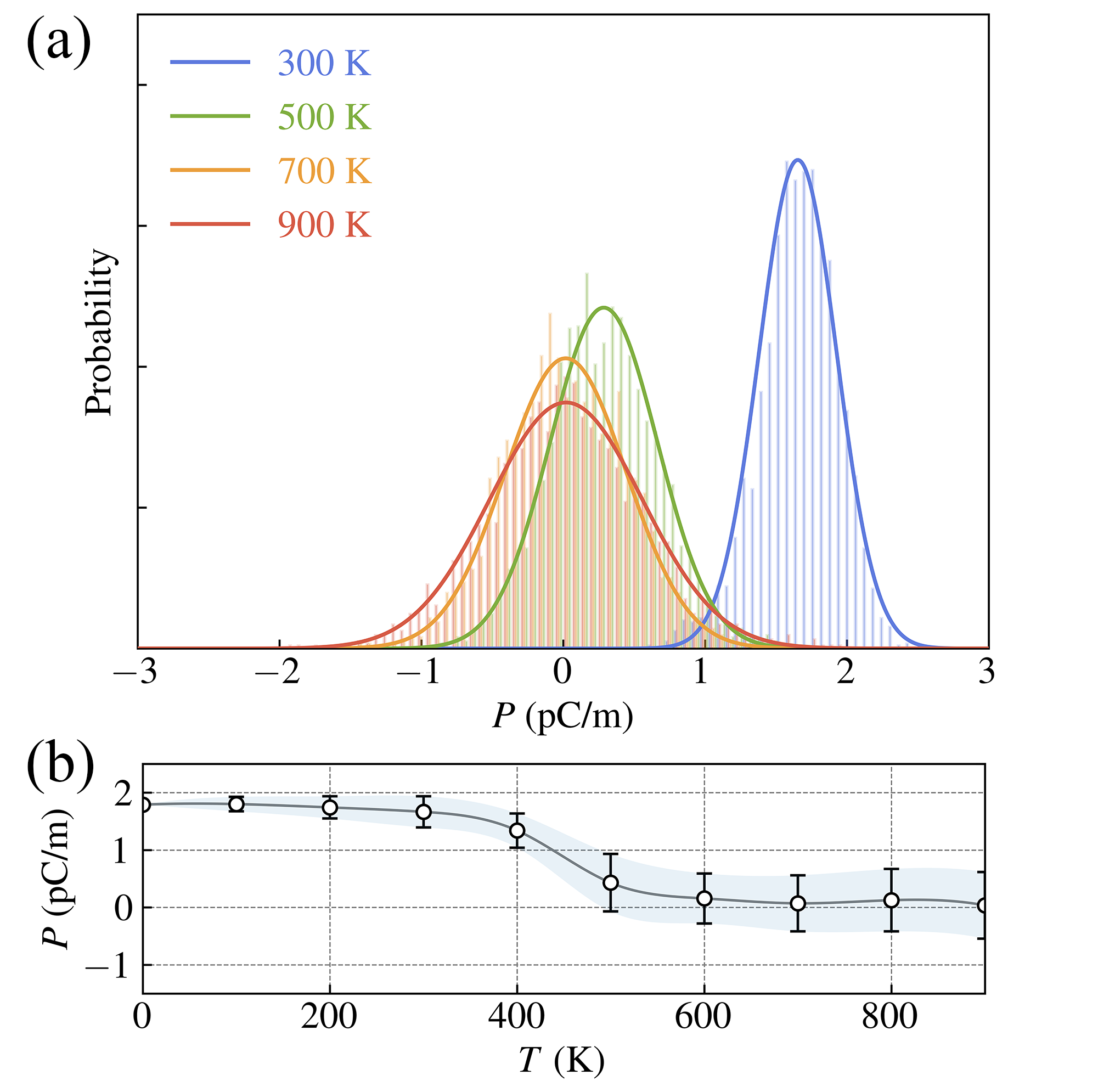}
 \centering
 \caption{(a) The probability of FE polarization follows a normal (Gaussian) distribution at 300, 500, 700, and 900 K. (b) The mean $\mu$ (represented by circles) and the standard deviation $\sigma$ (indicated by error bars) are shown for each temperature.
 }
 \label{MD}
 \end{figure}

\subsection*{Ferroelectric properties}
In addition to its 2D FM metallic behavior, the Cu(CrSe$_2$)$_2$ monolayer may also exhibit FE properties due to its space group $P3m1$ and lack of spatial inversion symmetry.
As seen in Fig.~\ref{struct}(a), the originally centrosymmetric bilayer of CrSe$_2$ is disrupted by the intercalation of Cu ions, resulting in an out-of-plane FE polarization~\cite{li_2021}.
In this study, the dipole moment is computed directly by integrating the charge density (including both electronic and ionic contributions) of the metallic thin film. The FE polarization is calculated by applying a dipole moment correction to evaluate the vertical polarization of the Cu(CrSe$_2$)$_2$ monolayer. This method has already been justified in the study of the similar FE metallic WTe$_2$ bilayer~\cite{WTe22018}.
Our calculations indicate that the monolayer exhibits an FE polarization of approximately 1.79 pC/m in the negative direction along the c-axis, representing a vertical downward FE polarization ($-$P).
In 2D metallic systems, electrons move freely within the basal plane but face constraints in the vertical direction. This characteristic might enable the persistence of out-of-plane polarization, which could be altered by a perpendicular electric field, as suggested for the 2D FE metal WTe$_2$~\cite{Fei2018}.

To confirm the stability of the FE phase in the Cu(CrSe$_2$)$_2$ monolayer, we conducted phonon calculations. The absence of imaginary frequencies over the Brillouin zone confirms the dynamic stability of the FE phase, as seen in Fig. S4 in SI. 
Moreover, the robustness of the FE state under ambient conditions is confirmed by our AIMD simulations, as shown in Fig. S5 in SI.
Note that the anti-FE configuration is unfavorable in energy more than 92 meV/f.u. compared with the FE state, as seen in Fig. S6 in SI.

To explore the stability of ferroelectricity under finite temperatures and to ascertain its FE phase transition temperature, we conducted AIMD simulations. 
We statistically analyze the thermal equilibrium states between 5 ps and 30 ps, as shown in Fig. S5, to calculate the probability of FE polarization.
The results indicate that the probability of FE polarization exhibits a singular peak and adheres to the normal (Gaussian) distribution at each temperature, represented as $P$ $\sim$ $\mathcal{N}$($\mu$,$\sigma^2$). Herein, $\mu$ represents the mean value, and $\sigma$ denotes the standard deviation value. 
As seen in Fig.~\ref{MD}, the material exhibits robust ferroelectricity above room temperature (even about 500 K). This result agrees well with the most recent experimental observation of the FE phase in monolayer CuCrSe$_2$ above room temperature~\cite{sun2024}.
Hence, Cu(CrSe$_2$)$_2$ monolayer could be an optimal 2D room-temperature FE material, in addition to its FM metallic behavior, and it would then be a new 2D multiferroic metal.

\begin{figure}[t]
  \centering
 \includegraphics[width=8.5cm]{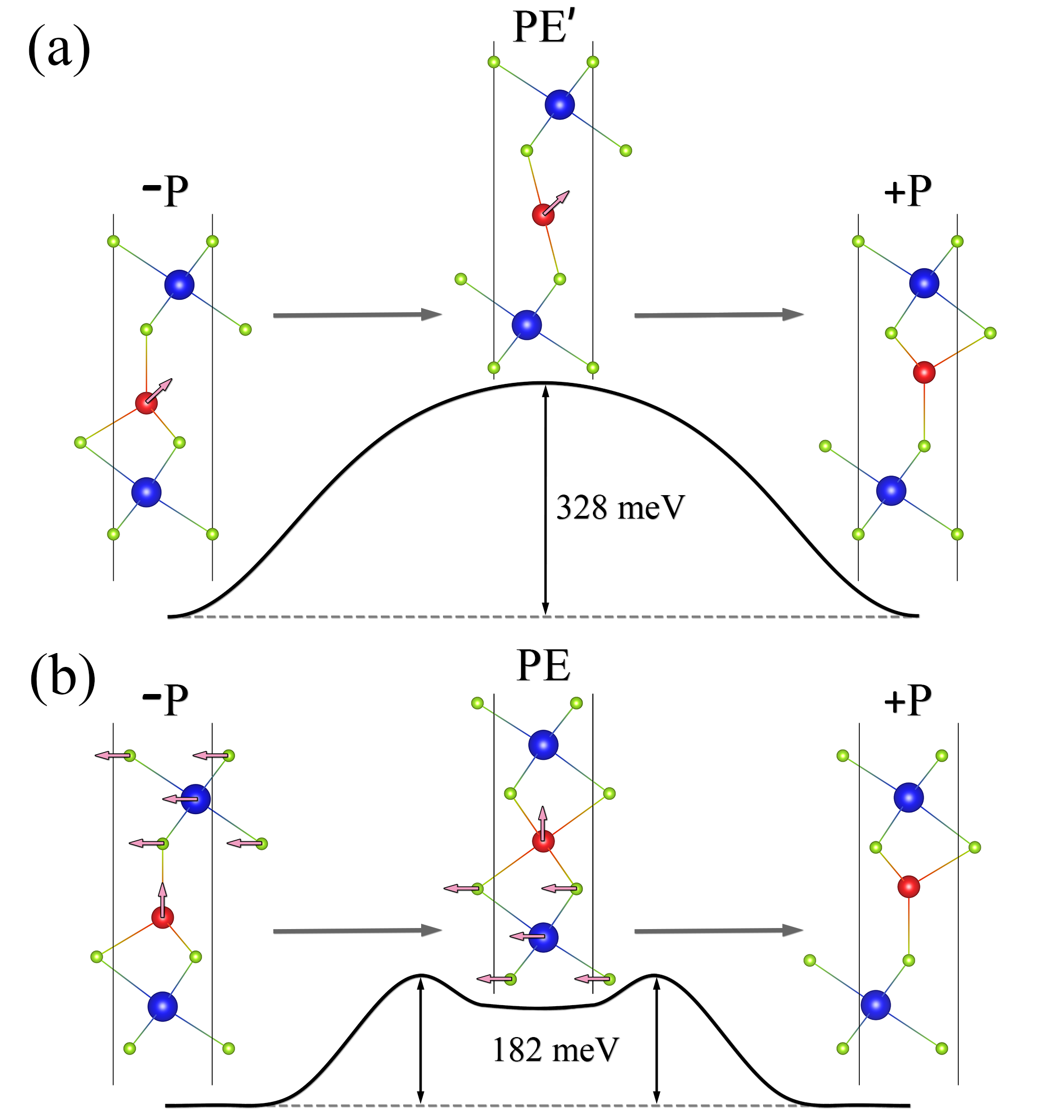}
 \centering
 \caption{FE switching pathways via the NEB method:
 (a) stable FE states with opposing polarizations ($-$P and $+$P) depicted on the left and right, respectively. The intermediate paraelectric state (PE$'$) in the middle arises solely from the displacement of the Cu ions. (b) Another intermediate paraelectric state (PE) shown in the middle, achieved by both the displacement of Cu ions and in-plane shifts of the CrSe$_2$ layers, where Cu forms an octahedral coordination with neighboring Se atoms. Pink arrows denote atomic motion directions during polarization reversal.
 }
 \label{PE-FE}
 \end{figure}

The reversal barrier of FE polarization is crucial for its practical applications. 
Here, we explore three possible reversal pathways.
In the Cu(CrSe$_2$)$_2$ structure, shifting the Cu ions can flip the state from a $-$P polarization to an energetically equivalent $+$P polarization, as illustrated in Fig.~\ref{PE-FE}. 
To elucidate the specific pathway for this switching in the Cu(CrSe$_2$)$_2$ monolayer, we utilized the NEB method~\cite{CINEB}. The results, depicted in Fig. ~\ref{PE-FE}(a), reveal that during the transition phase, Cu ions adopt a two-coordinated state, representing the paraelectric phase (PE$'$) as shown in Fig. ~\ref{PE-FE}(a). 
With a calculated switching barrier of 328 meV for each Cu ion, this large reversal barrier suggests that a high electric field would likely be required to reverse the polarization. 
Second, as a switching pathway via the anti-FE structure is possible~\cite{Huang2022}, we calculate it and find the switching barrier is 203 meV/f.u. Furthermore, the third pathway is explored and turns out to be most economic for an electric switch.
Here, the CrSe$_2$ layers exhibit in-plane shifts as indicated by the pink arrows, and each Cu ion is coordinated with six Se atoms to form the CuSe$_6$ octahedron as shown in Fig. ~\ref{PE-FE}(b). 
Our calculations indicate a much reduced switching barrier of 182 meV/f.u., well comparable with those of conventional 3D FE materials such as PbTiO$_3$ (200 meV/f.u.)~\cite{cohen1992}.
It is worth noting that this intermediate PE state is a local energy minimum, with its dynamic stability being corroborated also by phonon calculations, as seen in Fig. S4 in SI.

\subsection*{Magnetoelectric, doping, and strain effects}

\begin{figure}[t]
  \centering
 \includegraphics[width=8.5cm]{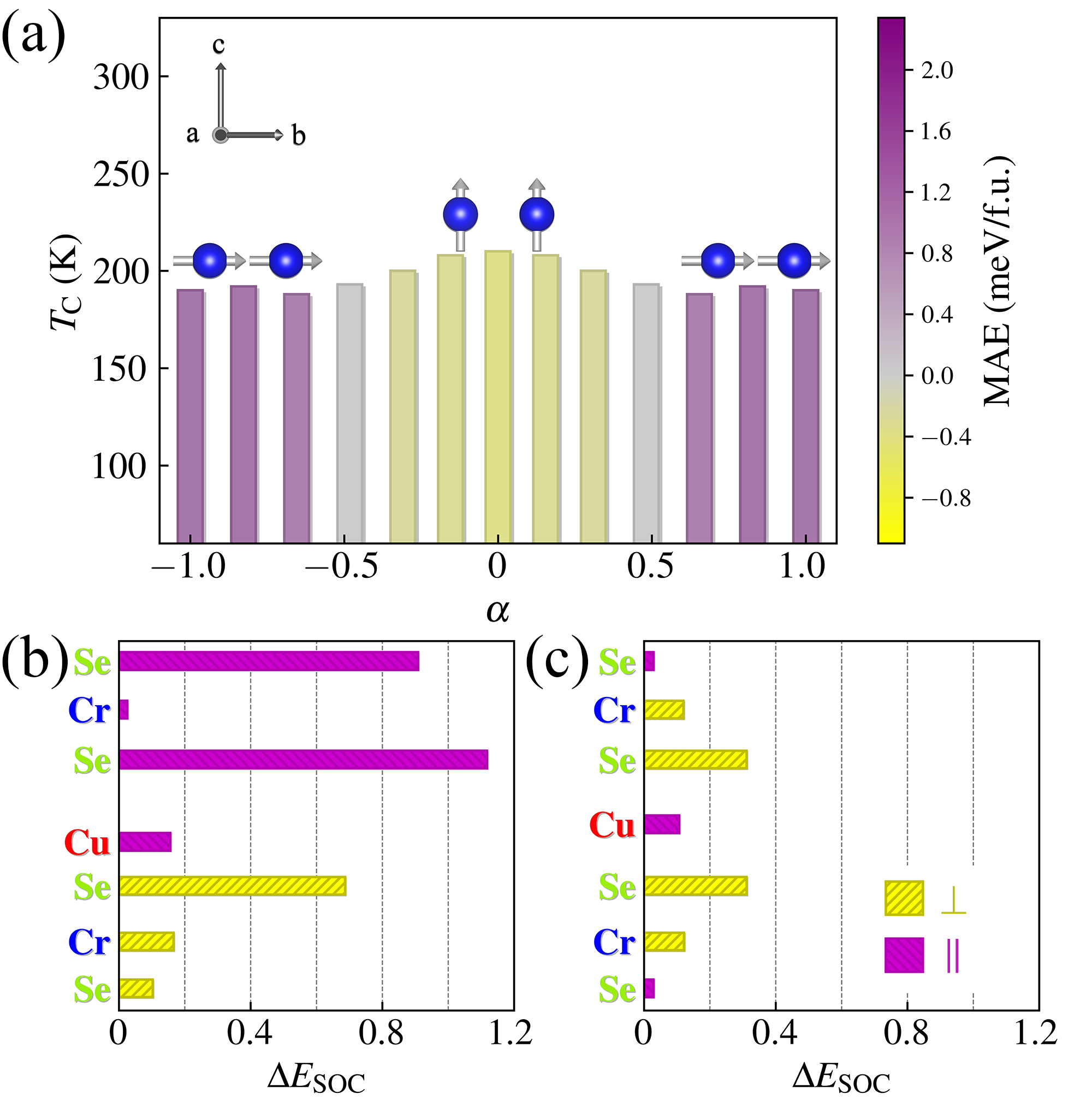}
 \centering
 \caption{(a) The variation of $T_{\rm C}$ and MAE with respect to the normalized amplitude of the polar distortion $\alpha$. The color bar represents the MAE. Yellow indicates that the easy magnetization axis is oriented out-of-plane, while purple signifies it is oriented in-plane.
The contribution to MAE by individual atomic layers for (b) $-$P structure and (c) PE structure: yellow ($\bot$) represents out-of-plane, while purple ($\parallel$) indicates in-plane.
 }
 \label{FEME}
 \end{figure}

Our above results suggest that the Cu(CrSe$_2$)$_2$ monolayer is not only a 2D FM metal but also FE, and thus it is added as a new case into the rare 2D multiferroic metals.
As seen in Fig. ~\ref{PE-FE}(b), although the PE phase has a higher energy than the FE phase, it is a local energy minimum. Thus, the PE state could stably exist during the FE switching. 
Next, we examine a potential magnetoelectric effect, namely, varying magnetism associated with the FE-PE transition driven by an electric field. Our results show that the MA's direction switches from the in-plane to out-of-plane during the FE-PE transition, as seen in Fig.~\ref{FEME}. Moreover, the $T_{\mathrm{C}}$ exhibits variations, peaking around the PE phase.
To understand the switch of MA axes, we computed the MAE contribution from each atomic layer. As observed from Figs. ~\ref{FEME}(b) and (c), the primary source of MAE in the FE state lies with the Se ion layers, being consistent with our above results. 
However, we observed significant changes in the MAE across the Se ion layers during the FE-PE transition. For example, the MAE of the second Se layer changes from the in-plane one of 1.12 meV/f.u. to the out-of-plane one of 0.31 meV/f.u. It is this change that plays a crucial role in the MA switch from in-plane to out-of-plane associated with the FE-PE transition.

\begin{figure}[t]
  \centering
 \includegraphics[width=8cm]{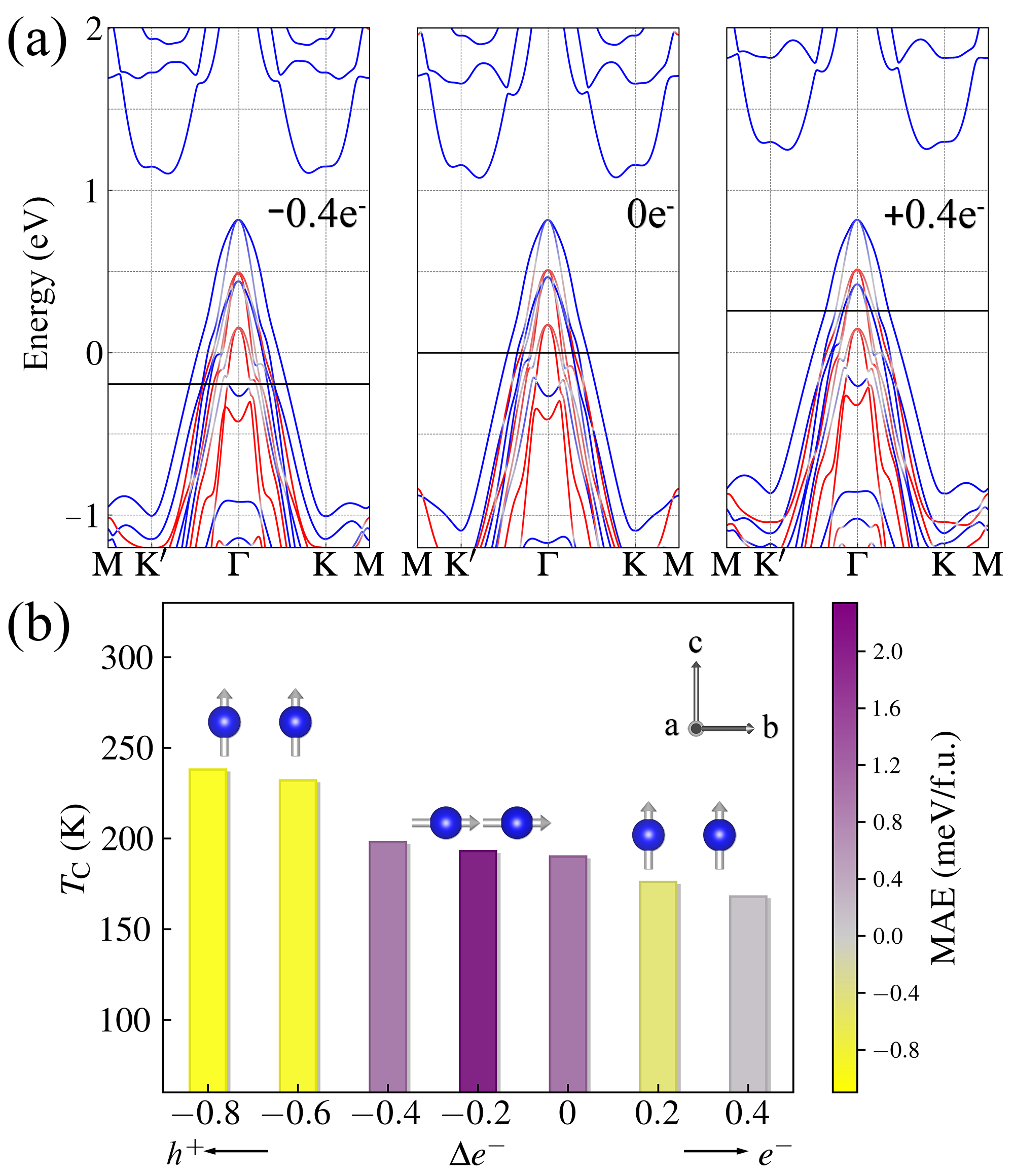}
 \centering
 \caption{(a) Band structures for doped Cu(CrSe$_2$)$_2$ monolayer. Red and blue lines represent down and up spin, respectively. The notation $\pm$$x$$e^-$ ($x$= 0, 0.4) indicates the number of doped electrons or holes per formula unit. The solid black line represents the Fermi level.
 (b) The variation of $T_{\rm C}$ and MAE with respect to the $\pm$$x$$e^-$. The color bar represents the MAE. Yellow indicates that the easy magnetization axis is oriented out-of-plane, while purple signifies it is oriented in-plane.
 }
 \label{doping}
 \end{figure}

Considering the material's FM itinerant nature and the manipulability of 2D materials, we further explore the control of its multiferroic properties through carrier dopings~\cite{deng2018,wu_2022}.
As illustrated in Figs.~\ref{struct}(c), a finite hole doping would increase the density of band states at the Fermi level and thus favors the itinerant FM, whereas a partial electron doping could have an opposite effect. In our calculations, the hole (electron) doping is simulated using the virtual crystal approximation (VCA)~\cite{Bellaiche2000}. 
As expected, hole doping enhances FM coupling, raising $T_{\mathrm{C}}$ from 190 K to 238 K with 0.8 holes per formula unit, as shown in Fig. ~\ref{doping}.
Notably, with the hole doping, the easy MA direction changes from the otherwise in-plane to the out-of-plane.

\begin{figure}[t]
  \centering
 \includegraphics[width=8.5cm]{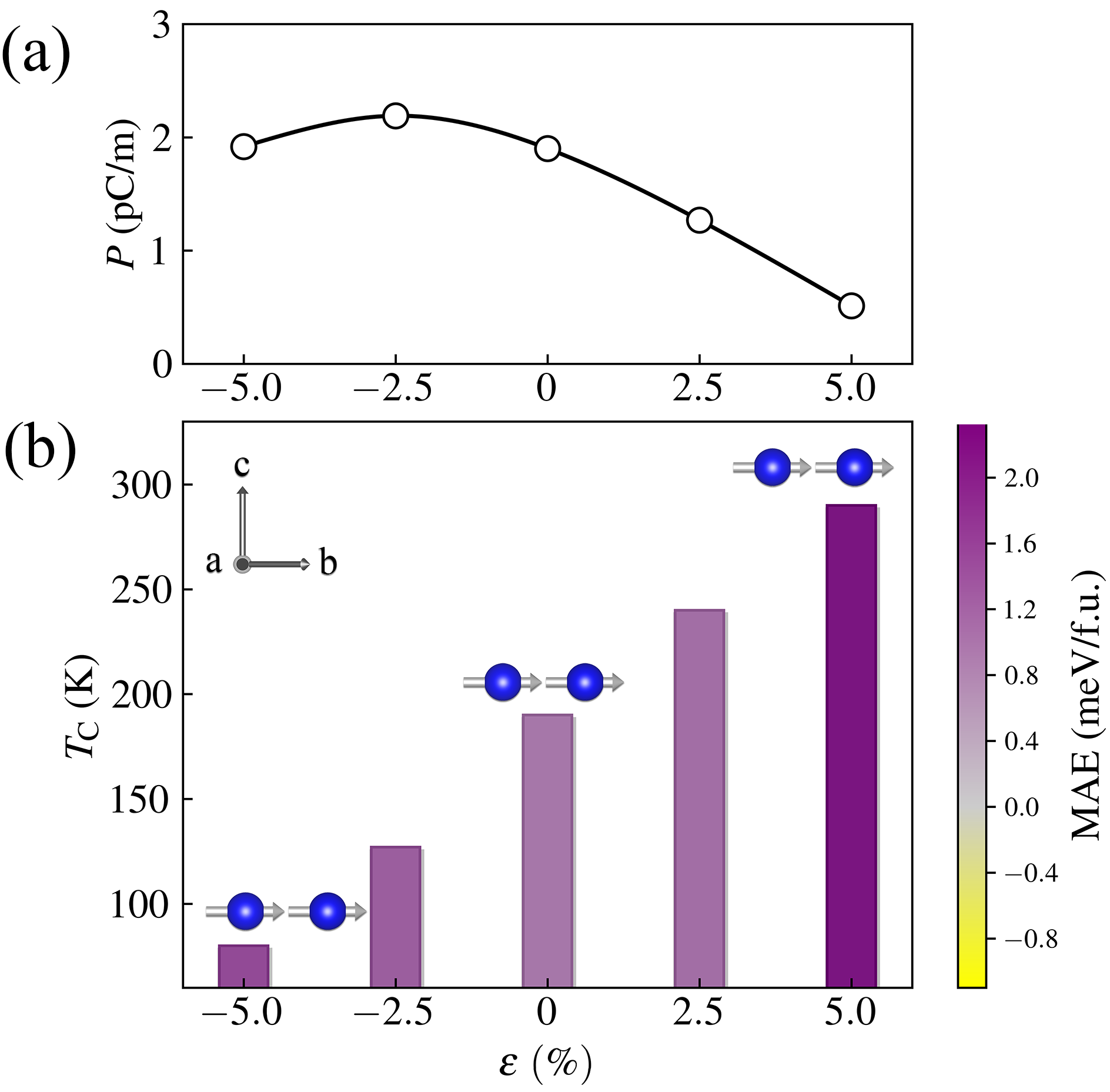}
 \centering
 \caption{(a) Variation curves of the FE polarization with respect to the biaxial strain $\varepsilon\%$. (b) The variation of $T_{\rm C}$ and MAE with respect to the $\varepsilon\%$. The color bar represents the MAE. yellow represents out-of-plane, while purple indicates in-plane.
 }
 \label{strain}
 \end{figure}

Strain is widely used to tune the properties of 2D materials~\cite{Cenker2022,Wang2020}.
We also examined the effects of biaxial strain (within $\pm$5\%) on the multiferroic Cu(CrSe$_2$)$_2$ monolayer. As seen in Fig.~\ref{strain}, under a tensile strain, the in-plane lattice constants increase while the thickness along the $c$-direction decreases, typically reducing the vertical FE values. Conversely, a finite compressive strain would enhance the FE polarization. We also observed that the FM $T_{\mathrm{C}}$ is increased up to 290 K by a tensile strain of 5\%.
All the results in this subsection suggest that the magnetoelectric properties of the rare 2D multiferroic metallic Cu(CrSe$_2$)$_2$ monolayer could be tuned by electric field, carrier doping, and lattice strain, thus highlighting its potential technological applications.

\section*{Summary}

We investigated the magnetoelectric properties and the relevant physics in the new 2D multiferroic metallic Cu(CrSe$_2$)$_2$ monolayer, using density functional calculations, Monte Carlo simulations, and $ab$ $initio$ molecular dynamics. Our results show that the monolayer is in the Cr$^{3+}$ $S$ = 3/2 and Cu$^{+}$ $S$ = 0 state. The ligand Se 4$p$ holes enhance the FM metallicity with the $T_{\rm C}$ about 190 K, and the Se $4p$-Cr $3d$ exchange determines the in-plane MA, in which the single ion anisotropy and shape MA are all much weaker or even negligible. Moreover, we predict that this monolayer retains the vertical FE polarization above room temperature. Therefore, Cu(CrSe$_2$)$_2$ monolayer seems to be a quite stable multiferroic metal. Furthermore, 
our results predict that the magnetoelectric properties could be tuned by electric field, carrier doping, and lattice strain: a MA switch from the in-plane to out-of-plane upon the FE-PE transition or hole doping, and hole doping or tensile strain enhanced $T_{\rm C}$ even up to room temperature. Thus, Cu(CrSe$_2$)$_2$ monolayer is a new and rare 2D multiferroic metal and could have promising spintronic applications.

\section*{Acknowledgements}
The authors thank Jinlian Lu at Yancheng Institute of Technology, and Zhe Wang and Binhua Zhang at Fudan University for fruitful discussions. 
This work was supported by National Natural Science Foundation of China (Grants No. 12104307, No. 12174062, and No. 12241402), and by Innovation Program for Quantum Science and Technology (2024ZD0300102). 

\bibliography{rsc}

\onecolumngrid
\newpage
\begin{appendix}
	\setcounter{figure}{0}
	\setcounter{table}{0}
	\renewcommand{\thefigure}{S\arabic{figure}}
	\renewcommand{\thetable}{S\arabic{table}}
	\renewcommand{\theequation}{S\arabic{equation}}
	\renewcommand{\tablename}{Table}
	\renewcommand{\figurename}{Fig.}

\titleformat*{\section}{\normalfont\Large\bfseries}
\section*{Supporting Information for "Multiferroic Metallic Monolayer Cu(CrSe$_2$)$_2$"}

\renewcommand\arraystretch{1.5}
\begin{table}[H]
  \centering
\caption{The calculated magnetic anisotropy energy MAE (meV/f.u.) for Cu(CrSe$_2$)$_2$ using different k-meshes and cut-off (eV) energies.}
\begin{tabular}{c@{\hskip5mm}c@{\hskip5mm}c@{\hskip5mm}c}
\hline\hline
k-mesh                &$E_\text{cut}$    & 010($b$)   &  001($c$)        \\ \hline
15$\times$15$\times$1 &500               &   0        &  1.084           \\
15$\times$15$\times$1 &550               &   0        &  1.092           \\
17$\times$17$\times$1 &500               &   0        &  1.124           \\
\hline\hline
 \end{tabular}
 \label{tb1}
\end{table}

\begin{figure}[H]
  \centering
\includegraphics[width=15cm]{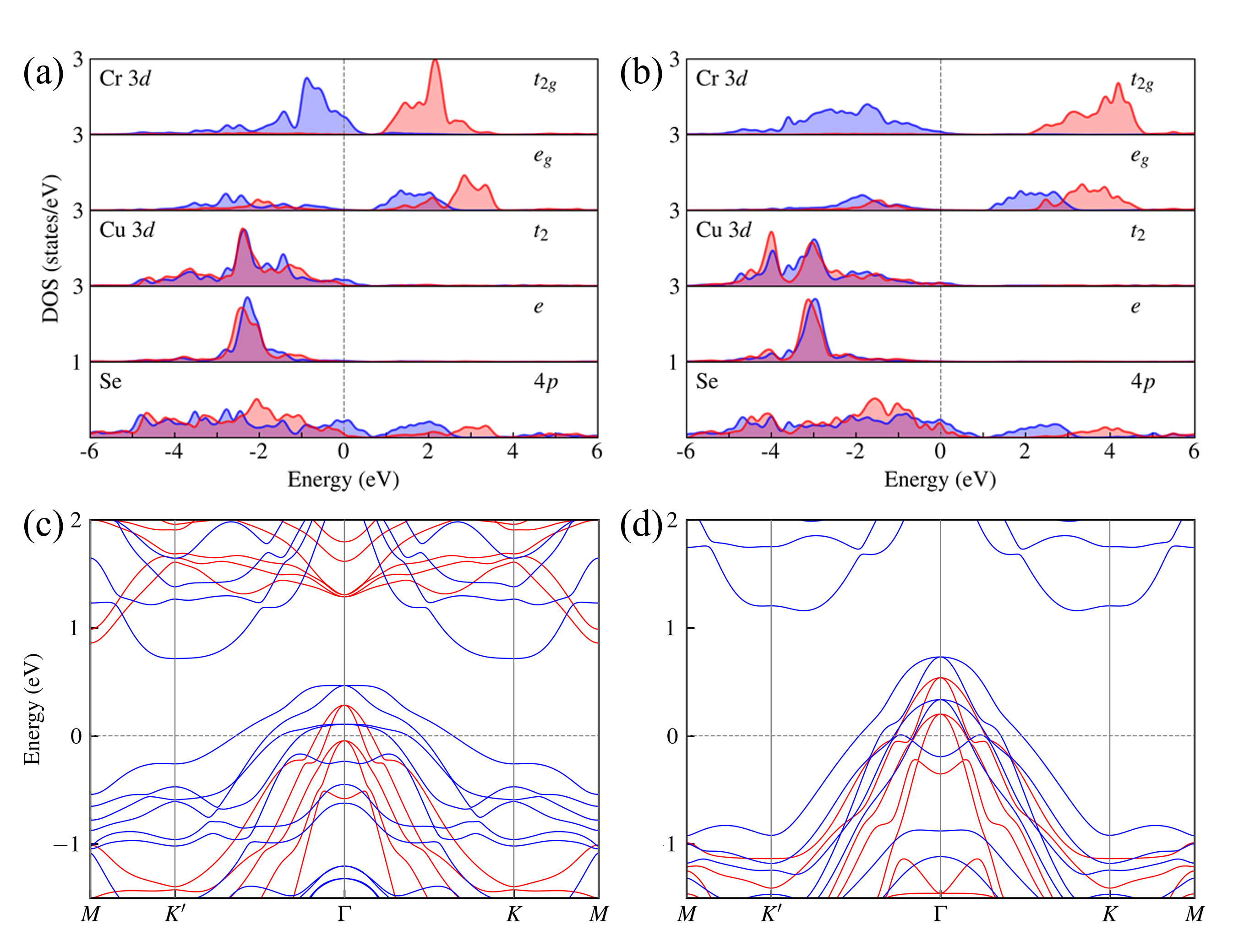}
\centering
 \caption{The Cr 3$d$, Cu 3$d$, Se 4$p$, and total density of states (DOS) are shown for (a) GGA and (b) GGA+$U$ calculations. The corresponding band structures are presented in (c) GGA and (d) GGA+$U$. The blue (red) curves stand for the up (down) spin channel. The Fermi level is set at zero energy.}
\label{FE_AFE}
\end{figure}

\renewcommand\arraystretch{1.5}
\begin{table}[H]
  \centering
\caption{The magnetic anisotropy energy (MAE, meV/f.u.), exchange interaction parameter (\( J \), meV), and Curie temperature (\( T_{\text{C}} \), K) by GGA + SOC and GGA + SOC + \( U \) for different \( U \) (eV) values.}
\begin{tabular}{c@{\hskip5mm}c@{\hskip5mm}c@{\hskip5mm}c@{\hskip5mm}c@{\hskip5mm}c@{\hskip5mm}c}
  \hline\hline
     &  &   MAE        &$J_{\text{top}}$ &$J_{\text{bottom}}$ &$J_{\text{inter}}$ &$T_{\text{C}}$\\ \hline
     \multicolumn{2}{c}{GGA}           &   $-$2.54    &$-$9.46          &$-$10.36             &$-$1.58            &360           \\
\hline\hline
$U_{\text{Cu}}$      & $U_{\text{Cr}}$ &   MAE        &$J_{\text{top}}$ &$J_{\text{bottom}}$ &$J_{\text{inter}}$ &$T_{\text{C}}$\\ \hline
6                    &   4             &   $-$1.08    &$-$5.70          &$-$6.55             &$-$0.18            &190           \\
6                    &   5             &   $-$1.40    &$-$5.89          &$-$7.09             &$-$0.03            &178           \\
7                    &   4             &   $-$0.95    &$-$5.73          &$-$6.53             &$-$0.19            &190           \\
7                    &   5             &   $-$1.30    &$-$5.91          &$-$7.04             &$-$0.04            &178           \\
\hline\hline
 \end{tabular}
 \label{tb2}
\end{table}
 \begin{figure}[H]
  \centering
 \includegraphics[width=8cm]{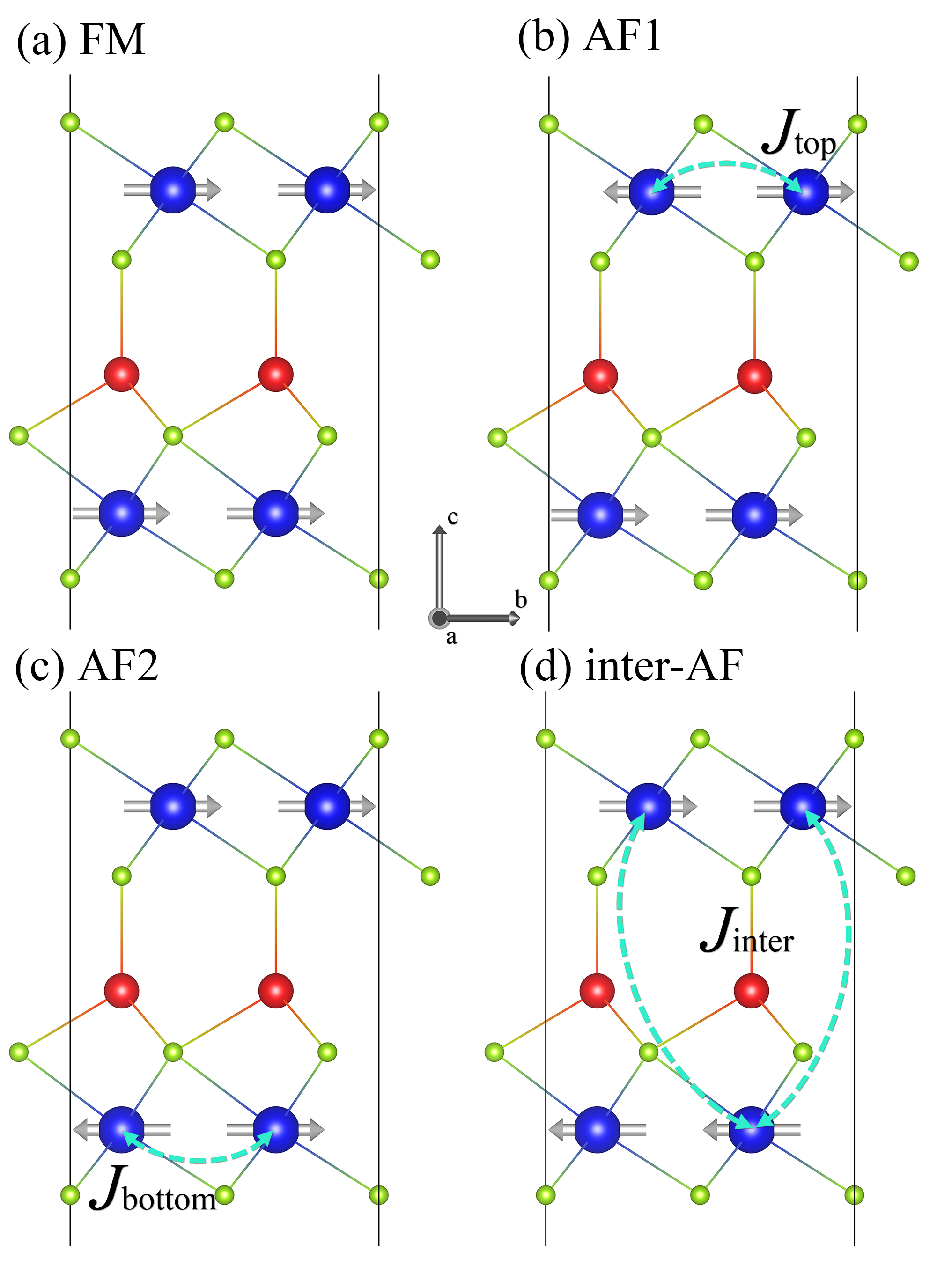}
 \centering
 \caption{The four magnetic structures of Cu(CrSe$_2$)$_2$ marked with three exchange parameters.
 }
 \label{states}
 \end{figure}
\renewcommand\arraystretch{1.5}
\begin{table}[H]
  \centering
\caption{Relative total energies $\Delta$\textit{E} (meV/f.u.) and local spin moments ($\mu_{\rm B}$) for the Cu(CrSe$_2$)$_2$ monolayer. The derived three exchange parameters are $J_{\rm {top}}$
=--5.70 meV and $J_{\rm {bottom}}$ =--6.55 meV, and the interlayer $J_{\rm {inter}}$ =--0.18 meV. The $T_{\rm C}$ of 190 K is estimated by MC simulation.}
\begin{tabular}{c@{\hskip5mm}c@{\hskip5mm}c@{\hskip5mm}c@{\hskip5mm}c@{\hskip5mm}c}
\hline\hline
state             &Energy    & $M_{\text{Cr}_{\text{top}}}$  & $M_{\text{Cr}_{\text{bottom}}}$  &  $M_{\text{Cu}}$\\ \hline
FM                &0         &  3.04                         &   3.12                           &  $-$0.05        \\
AF1           &60.17     &  $\pm$2.93                    &   3.12                           &  $-$0.03         \\
AF2       &52.47     &  3.05                         &   $\pm$2.98                      &  $-$0.01         \\
inter-AF          &2.38      &  3.05                         &   $-$3.12                        &  0.01         \\
\hline\hline
 \end{tabular}
 \label{tb3}
\end{table}
\renewcommand\arraystretch{1.5}
\begin{table}[H]
  \centering
\caption{The relative total energy (meV/f.u.) for the different states is calculated using the four-state method, employing a 3$\times$4 supercell to calculate the Dzyaloshinskii-Moriya (DM) parameters (meV) of the top layer CrSe$_2$. This involves a pair of nearest Cr ions along the b-axis in the top layer.}
\begin{tabular}{c@{\hskip5mm}c@{\hskip5mm}|c@{\hskip5mm}|c@{\hskip5mm}|c@{\hskip5mm}}
\hline\hline
 $\tau_1$ & $\tau_2$ & $(0,\tau_1 S,0);\ (0,0,\tau_2 S)$ & $(0,0,\tau_1 S);\ (\tau_2 S,0,0)$ & $(\tau_1 S,0, 0);\ (0,\tau_2 S,0)$ \\ \hline
 $+$      & $+$      & $0$                              & $0$                               & $0$                                \\
 $-$      & $+$      & $0.247$                           & $0.457$                           & $124.801$                          \\
 $+$      & $-$      & $124.045$                         & $-0.169$                          & $0.061$                            \\
 $-$      & $-$      & $125.836$                         & $-0.714$                          & $124.938$                          \\    
          &          & $D_x=0.09$                        & $D_y=-0.08$                       & $D_z=0.09$                         \\
\hline\hline
 \end{tabular}
 \label{tb5}
\end{table}
\renewcommand\arraystretch{1.5}
\begin{table}[H]
  \centering
\caption{The relative total energy (meV/f.u.) for the different states is calculated using the four-state method, employing a 3$\times$4 supercell to calculate the DM parameters (meV) of the bottom layer CrSe$_2$. This involves a pair of nearest Cr ions along the b-axis in the bottom layer.}
\begin{tabular}{c@{\hskip5mm}c@{\hskip5mm}|c@{\hskip5mm}|c@{\hskip5mm}|c@{\hskip5mm}}
\hline\hline
 $\tau_1$ & $\tau_2$ & $(0,\tau_1 S,0);\ (0,0,\tau_2 S)$ & $(0,0,\tau_1 S);\ (\tau_2 S,0,0)$ & $(\tau_1 S,0, 0);\ (0,\tau_2 S,0)$ \\ \hline
 $+$      & $+$      & $0$                               & $0$                               & $0$                                \\
 $-$      & $+$      & $0.487$                           & $-1.012$                          & $121.444$                          \\
 $+$      & $-$      & $121.441$                         & $0.563$                           & $0.319$                            \\
 $-$      & $-$      & $122.763$                         & $-1.167$                          & $122.580$                          \\   
          &          & $D_x=0.17$                        & $D_y=-0.11$                       & $D_z=0$                         \\
\hline\hline
 \end{tabular}
 \label{tb6}
\end{table}

\begin{figure}[H]
  \centering
\includegraphics[width=18cm]{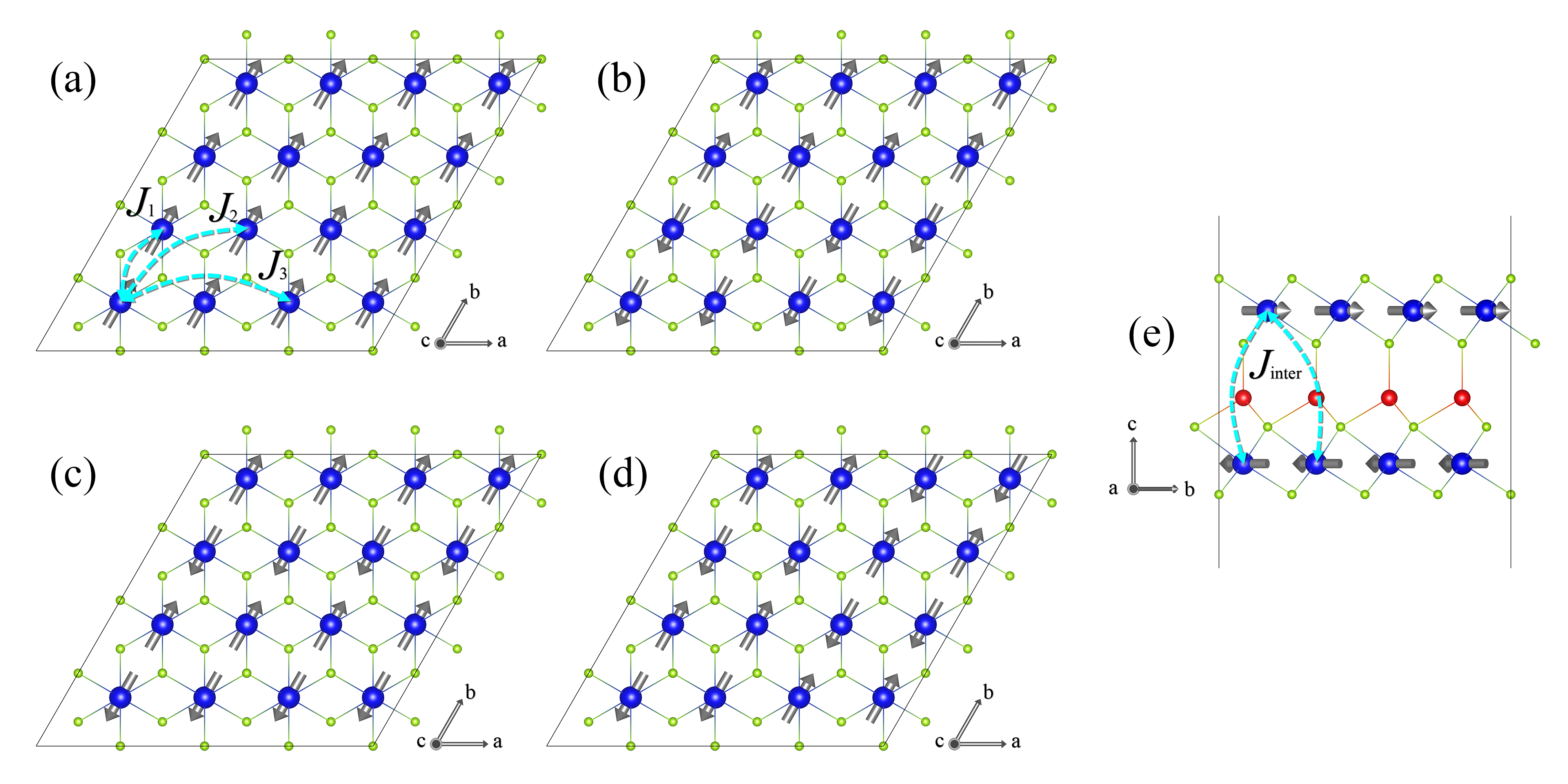}
\centering
 \caption{The five magnetic configurations of Cu(CrSe$_2$)$_2$ are utilized to estimate the second-nearest neighbor $J_2$, and the third-nearest neighbor $J_3$ within the CrSe$_2$ layer.
}
\label{states}
\end{figure}

\begin{equation}\begin{split}
E_{\text{FM},\text{FM}}&=\left[3(J_{\text{top}1}+J_{\text{top}2}+J_{\text{top}3})
+3(J_{\text{bottom}1}+J_{\text{bottom}2}+J_{\text{bottom}3}
)+3J_{\text{inter}}\right]S^2\\
E_{\text{AF1},\text{FM}}&=\left[(J_{\text{top}1}-J_{\text{top}2}-J_{\text{top}3})
+3(J_{\text{bottom}1}+J_{\text{bottom}2}+J_{\text{bottom}3}
)\right]S^2\\
E_{\text{AF2},\text{FM}}&=\left[(-J_{\text{top}1}-J_{\text{top}2}+J_{\text{top}3})
+3(J_{\text{bottom}1}+J_{\text{bottom}2}+J_{\text{bottom}3}
)\right]S^2\\
E_{\text{AF3},\text{FM}}&=\left[(-J_{\text{top}1}+J_{\text{top}2}-J_{\text{top}3})
+3(J_{\text{bottom}1}+J_{\text{bottom}2}+J_{\text{bottom}3}
)\right]S^2\\
E_{\text{FM},\text{FM}'}&=\left[3(J_{\text{top}1}+J_{\text{top}2}+J_{\text{top}3})
+3(J_{\text{bottom}1}+J_{\text{bottom}2}+J_{\text{bottom}3}
)-3J_{\text{inter}}\right]S^2\\
E_{\text{FM},\text{AF1}}&=\left[3(J_{\text{top}1}+J_{\text{top}2}+J_{\text{top}3})
+(J_{\text{bottom}1}-J_{\text{bottom}2}-J_{\text{bottom}3}
)\right]S^2\\
E_{\text{FM},\text{AF2}}&=\left[3(J_{\text{top}1}+J_{\text{top}2}+J_{\text{top}3})
+(-J_{\text{bottom}1}-J_{\text{bottom}2}+J_{\text{bottom}3}
)\right]S^2\\
E_{\text{FM},\text{AF3}}&=\left[3(J_{\text{top}1}+J_{\text{top}2}+J_{\text{top}3})
+(-J_{\text{bottom}1}+J_{\text{bottom}2}-J_{\text{bottom}3}
)\right]S^2\\
\end{split}\end{equation}

\renewcommand\arraystretch{1.5}
\begin{table}[H]
  \centering
\caption{The derived exchange parameters (meV) up to the third-nearest neighbors within the CrSe$_2$ layer. The $T_{\rm C}$ of 185 K is estimated by MC simulation.}
\begin{tabular}{c@{\hskip5mm}c@{\hskip5mm}c@{\hskip5mm}c}
\hline\hline
type                &$J_1$                & $J_2$        &  $J_3$           \\ \hline
top layer           &$-$4.59               &$-$0.73        &0.15              \\
bottom layer        &$-$6.23               &$-$0.50        &0.56              \\
inter-layer         &$-$0.26               &   $/$         &  $/$              \\
\hline\hline
 \end{tabular}
 \label{tb4}
\end{table}

\begin{figure}[H]
  \centering
\includegraphics[width=18cm]{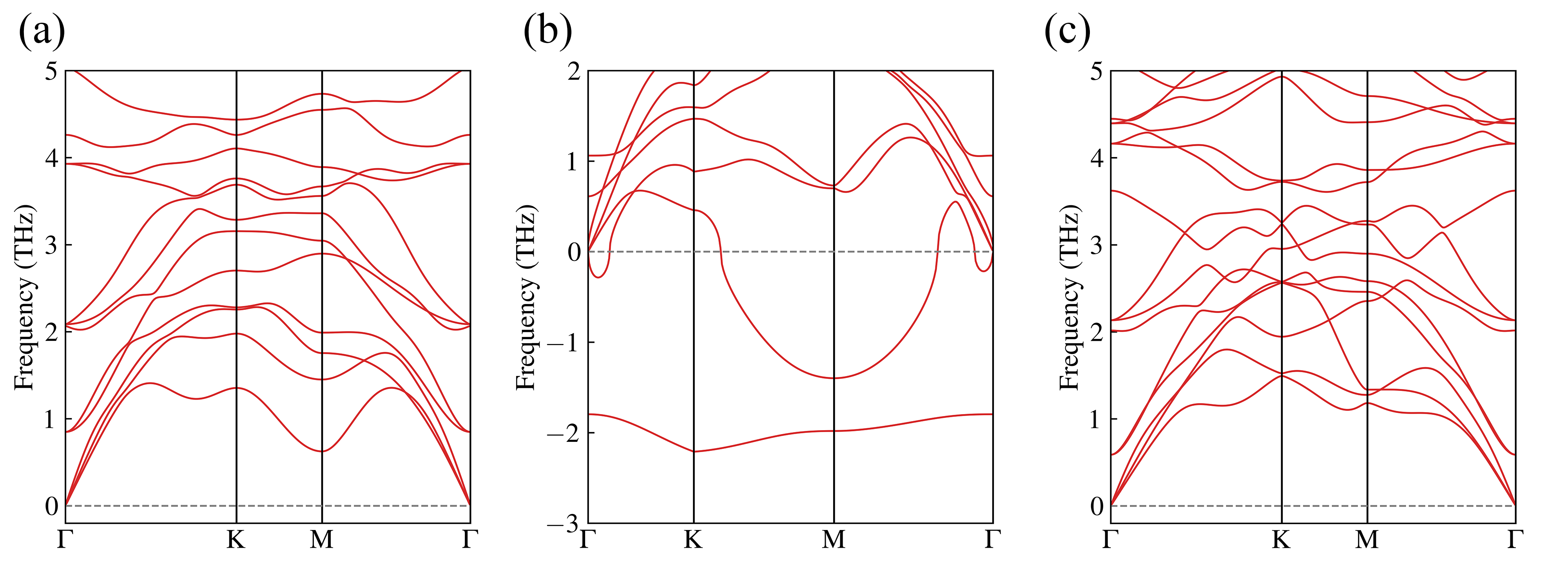}
\centering
 \caption{The phonon spectrum of (a) FE ($\pm$P) (b) PE’, and (c) PE states, see also Fig. 3 in the main text.}
\label{phonon}
\end{figure}

\begin{figure}[H]
  \centering
\includegraphics[width=9cm]{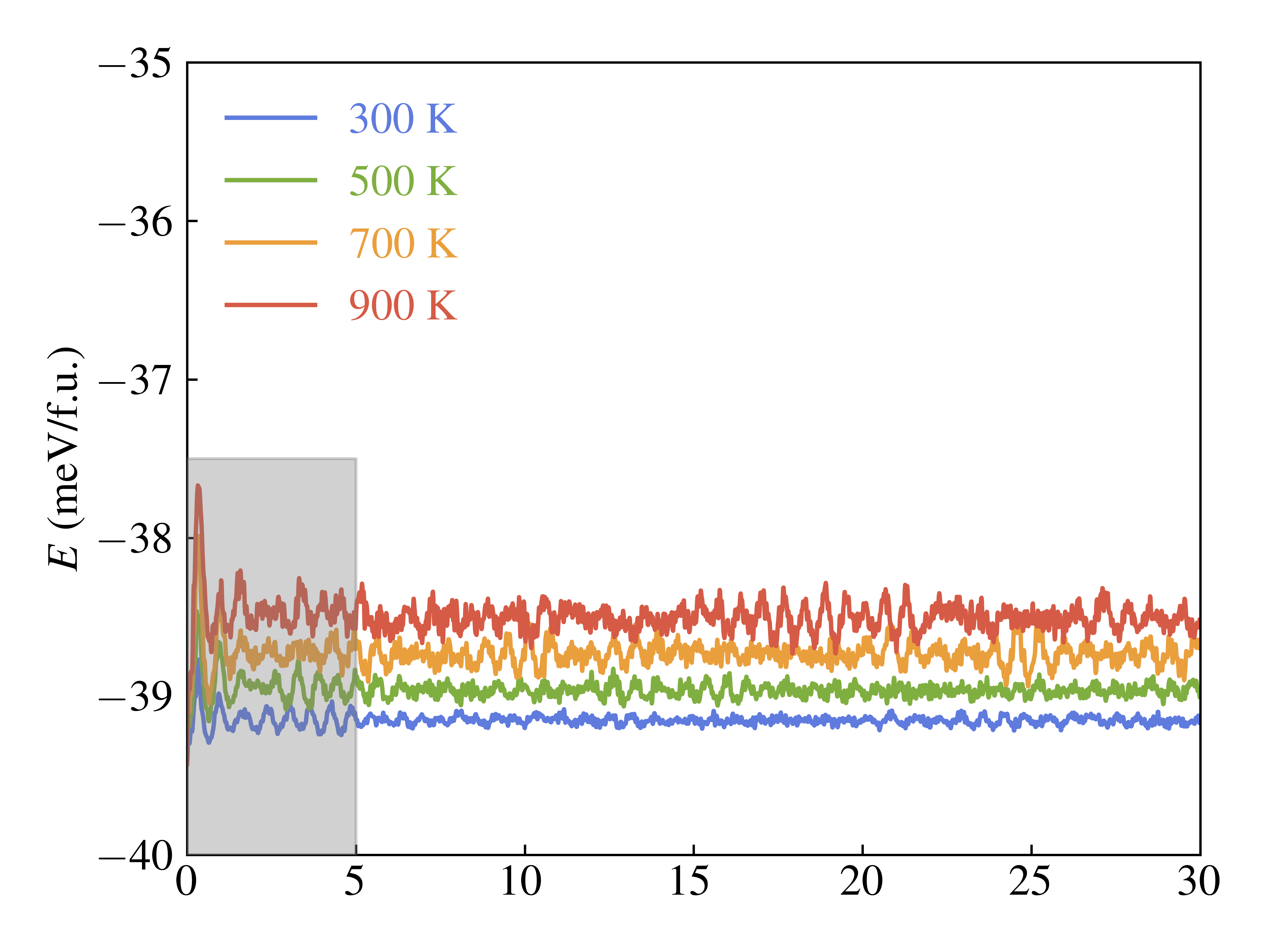}
\centering
 \caption{The total-energy evolution in the ab initio molecular dynamics simulation.}
\label{md_ene}
\end{figure}

\begin{figure}[H]
  \centering
\includegraphics[width=18cm]{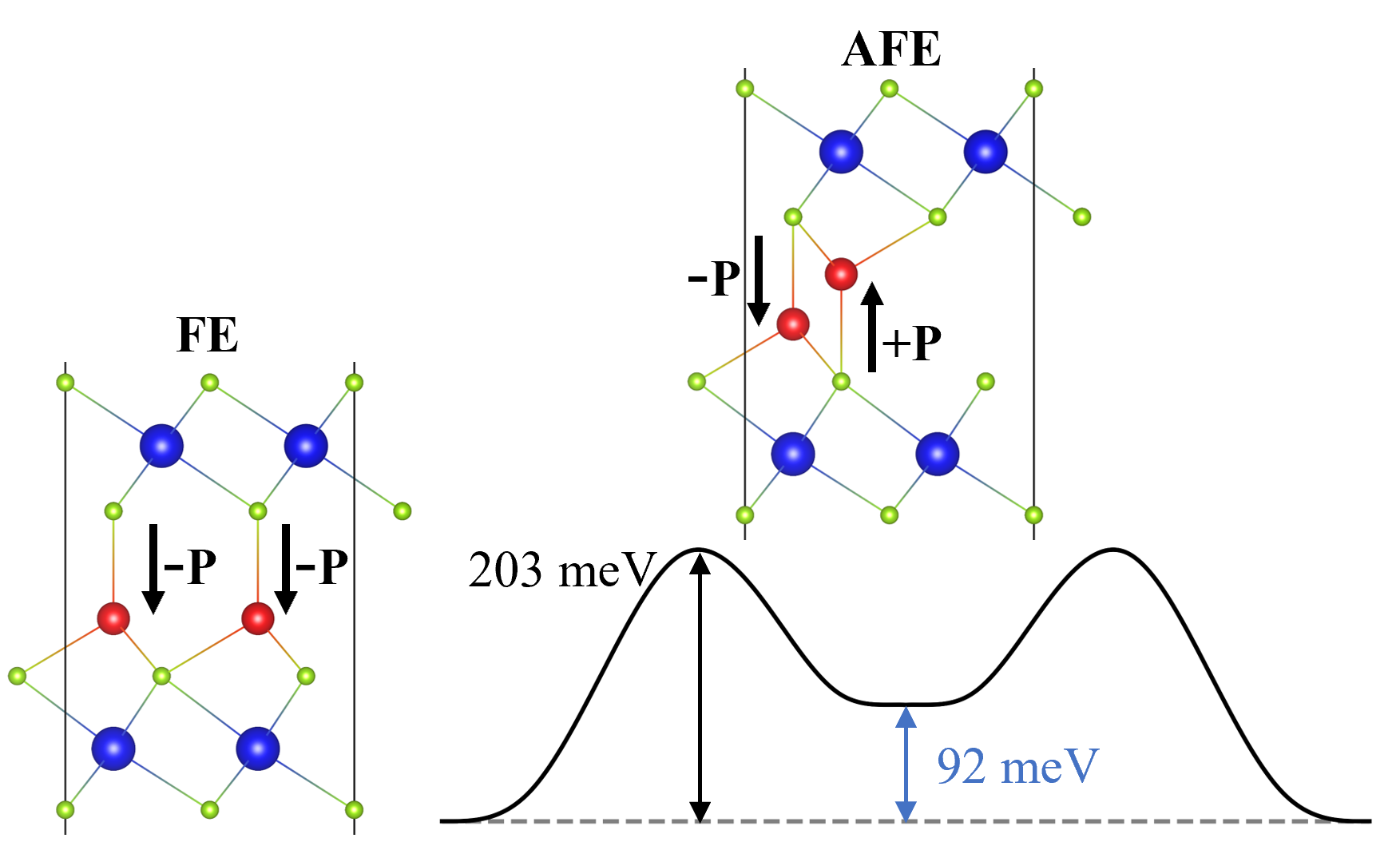}
\centering
 \caption{The FE and anti-FE structure. The anti-FE one is less stable by 92 meV/f.u.}
\label{FE_AFE}
\end{figure}

\end{appendix}
\end{document}